# TENSOR INTERACTIONS IN MEAN-FIELD APPROACHES[*]

J. DOBACZEWSKI

*Institute of Theoretical Physics, Warsaw University*
*ul. Hoża 69, 00-681 Warsaw, Poland*
*E-mail: Jacek.Dobaczewski@fuw.edu.pl*

Basic properties of the nuclear tensor mean fields are reviewed, and their role in changing the shell structure and masses of nuclei is analyzed within the spherical Hartree-Fock-Bogolyubov approach.

## 1. Introduction

Modern formulation of the nuclear mean-field theory is based on the energy density formalism,[1] which has been over the years developed for electronic systems. According to the formal Hohenberg-Kohn[2] and Kohn-Sham theorems,[3] exact ground-state energies of many-fermion systems can be obtained by minimizing certain exact functional of one-body density. These theorems do not provide any method to construct the exact functional in a systematic way; nevertheless one can build phenomenological functionals and test their performance against experimental data. Such an approach is also consistent with the ideas of the effective field theory, whereupon properties of composite objects at low energies can be described by Lagrangians which include high-energy dynamics in the form of the appropriate series of contact terms.

Within the energy density formalism, one treats the nucleus as a single composite object described by a set of one-body densities. At low energies, when the densities are varying slowly in the nuclear interior and then go smoothly to zero at the nuclear surface, one can consider only local densities that are built of the one-body density matrix and its derivatives up to the second-order. Systematics construction of the most general energy density functional (EDF) consistent with symmetries is then possible,[4] and gives

---

[*] Talk presented at the 3rd ANL/MSU/INT/JINA RIA Theory Workshop, Argonne, 4-7 April 2006

a generalization of the extremely successful approach based on the Skyrme effective interaction.[5]

The origins of the spin-orbit (SO) splitting in nuclei can be attributed to the bare two-body SO and tensor interactions,[6] which contribute differently to the spin-saturated (SS) and spin-unsaturated (SUS) nuclei. An alternative explanation is also sought in the relativistic mean-field theories with meson couplings[7], where no distinction between the SS and SUS systems is obtained.

The Skyrme interaction was introduced into nuclear physics more then 30 years ago,[8] and shortly after it was supplemented by tensor forces.[9,10] However, after these ground-breaking studies, in most of the subsequent applications the tensor forces were not taken into account. Moreover, within many Skyrme-force parameterizations constructed to date, the tensor terms in the EDF that were coming from the central force, were quite arbitrarily set equal to zero.[5]

In the present paper, I discuss the form of the tensor terms in the EDF and their influence on the single-particle energies and nuclear masses. In Sec. 2, I recall the recent experimental evidence on the changes of shell structure in neutron-rich $Z \approx 20$ nuclei.[11,12] This is only one of several such examples recently identified in light nuclei and interpreted within the shell-model by introducing tensor interactions.[13,14] Properties of the tensor terms in the EDF are discussed in Sec. 3, and in Sec. 4, I present results of calculations for single-particle energies and masses obtained with tensor terms included in the mean-field approach.

## 2. Shell structure of neutron-rich $Z \approx 20$ nuclei

In a series of recent experiments performed at the Argonne National Laboratory with Gammasphere[11] and National Superconducting Cyclotron Laboratory,[12] properties of low-lying collective states of even-even neutron-rich titanium isotopes have been measured. As illustrated in Fig. 1, the data reveal the presence of a closed $N=32$ subshell, in addition to the standard $N=28$ shell present in heavier elements. Indeed, both $^{50}$Ti and $^{54}$Ti show the increased $2^+$ energies, and decreased BE2 values, as compared to their neighbours.

In order to explain such a change of the shell structure, the single-particle neutron $\nu f_{5/2}$ orbital must be shifted up, which leaves a gap between the spin-orbit-split $\nu p_{3/2}$ and $\nu p_{1/2}$ orbitals, and creates a subshell closure for four particles occupying $\nu p_{3/2}$. The shell-model calculations,[15] performed for the single-particle orbitals shifted in this way, confirm the pattern shown in Fig. 1. The origins of the shift are attributed to the decreased monopole interaction energy between the proton $\pi f_{7/2}$ and neutron $\nu f_{5/2}$ orbitals, which occurs when protons are removed from $\pi f_{7/2}$. The source of such a monopole interaction is in turn attributed to the shell-model tensor interaction between these orbitals.

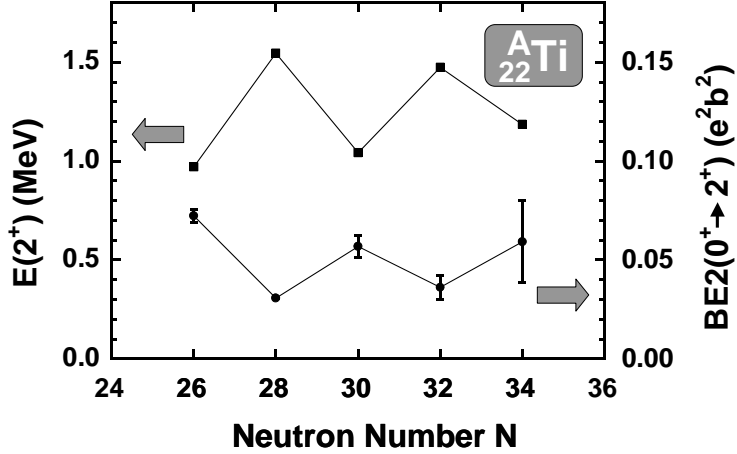

Figure 1. Energies of the first excited $2^+$ states (squares, left scale) and reduced transition probabilities BE2($0^+ \rightarrow 2^+$) (circles, right scale), measured in neutron-rich Ti isotopes.[11,12]

Positions of single-particle levels can be best studied within the mean-field approximation, in which they are basic dynamic characteristics of the system, resulting from the two-body interactions being averaged with particle densities of occupied states. Therefore, in this paper I evaluate the single-particle energies by applying the mean-field methods to tensor interactions.

## 3. Tensor densities in the energy density functional

### 3.1. *Spin-orbit and tensor forces*

Momentum-dependent two-body SO[8] and tensor[9,10] interactions have the form

$$
\begin{aligned}
\hat{V}_{SO} &= iW_0 \hat{\boldsymbol{S}} \cdot \left[ \hat{\boldsymbol{k}}' \times \hat{\boldsymbol{k}} \right], \\
\hat{V}_{Te} &= \tfrac{1}{2} t_e \left[ \hat{\boldsymbol{k}}' \cdot \hat{\mathsf{S}} \cdot \hat{\boldsymbol{k}}' + \hat{\boldsymbol{k}} \cdot \hat{\mathsf{S}} \cdot \hat{\boldsymbol{k}} \right], \\
\hat{V}_{To} &= t_o \hat{\boldsymbol{k}}' \cdot \hat{\mathsf{S}} \cdot \hat{\boldsymbol{k}},
\end{aligned} \tag{1}
$$

where the vector and tensor spin operators read

$$\hat{\boldsymbol{S}} = \boldsymbol{\sigma}_1 + \boldsymbol{\sigma}_2,$$
$$\hat{\mathsf{S}}^{ij} = \frac{3}{2}\Big[\boldsymbol{\sigma}_1^i \boldsymbol{\sigma}_2^j + \boldsymbol{\sigma}_1^j \boldsymbol{\sigma}_2^i\Big] - \delta^{ij}\boldsymbol{\sigma}_1 \cdot \boldsymbol{\sigma}_2. \tag{2}$$

When averaged with one-body density matrices, these interactions contribute to the following terms in the EDF (see Refs.[16,4] for derivations),

$$\mathcal{H}_{SO} = \frac{1}{4}W_0\Big[3J_0\frac{d\rho_0}{dr} + J_1\frac{d\rho_1}{dr}\Big],$$
$$\mathcal{H}_T = \frac{5}{8}\Big[t_e J_n J_p + t_o(J_0^2 - J_n J_p)\Big], \tag{3}$$

where the conservation of time-reversal and spherical symmetries was assumed. Here, $\rho_t$ and $J_t$ are the neutron, proton, isoscalar, and isovector particle and SO densities[8,16,4] for $t=n$, $p$, 0, and 1, respectively.

Apart from the contribution of the SO energy density to the central potential, variation of the SO and tensor terms with respect to the densities yields the one-body SO potential for neutrons ($t=n$) and protons ($t=p$),

$$W_t^{SO} = \frac{1}{2r}\Bigg[W_0\Big(\frac{d\rho_0}{dr} + \frac{d\rho_t}{dr}\Big) + \frac{5}{8}\Big((t_e + t_o)J_0 - (t_e - t_o)J_t\Big)\Bigg]\hat{\boldsymbol{L}} \cdot \hat{\boldsymbol{S}}. \tag{4}$$

Hence, it is clear that the only effect of including the tensor interaction is a modification of the SO splitting of the single-particle levels, and that, from the point of view of one-body properties, tensor interactions act very similarly to the two-body SO interactions. However, the latter ones induce the SO splitting that is only weakly depending on the shell filling. This is so because the corresponding form-factor in Eq. (4) is given by the radial derivatives of densities. On the other hand, the SO splitting induced by the tensor forces depends strongly on the shell filling, because its form-factor is given by the SO densities $J(r)$. Indeed, when only one of the SO partners is occupied (SUS system), the SO density is large, and when both partners are occupied (SS system), the SO density is small, see Sec. 4 for numerical examples.

### 3.2.  *Spin-orbit and tensor energy densities*

Within the energy-density approach, one does not relate the EDF to an average of the two-body force, but one postulates the EDF based on symmetry conditions only. Then, the most general EDF, depending on the spin-current densities, reads[4]

$$\mathcal{H}_{SO} = \sum_{t=0,1} C_t^{\nabla J}\rho_t \boldsymbol{\nabla} \cdot \boldsymbol{J}_t,$$
$$\mathcal{H}_T = \sum_{t=0,1}\Big(C_t^{J0}\underline{J}_t^2 + C_t^{J1}\boldsymbol{J}_t^2 + C_t^{J2}\underline{\underline{J}}_t^2\Big), \tag{5}$$

where the spherical-symmetry condition has been released. The standard pseudoscalar $\underline{J}_t$, vector $\boldsymbol{J}_t$, and pseudotensor $\underline{\mathsf{J}}_t$ parts of the spin-current density,[16]

$$J_{abt} = \tfrac{1}{2i}\Big[(\boldsymbol{\nabla} - \boldsymbol{\nabla}')_a s_{bt}(\boldsymbol{r}, \boldsymbol{r}')\Big]_{\boldsymbol{r}=\boldsymbol{r}'}, \qquad (6)$$

are defined as

$$\begin{aligned}
\underline{J}_t &= \textstyle\sum_{a=x,y,z} J_{aat},\\
\boldsymbol{J}_{at} &= \textstyle\sum_{b,c=x,y,z} \epsilon_{abc} J_{bct},\\
\underline{\mathsf{J}}_{abt} &= \tfrac{1}{2} J_{abt} + \tfrac{1}{2} J_{bat} - \tfrac{1}{3} \underline{J}_t \delta_{ab}.
\end{aligned} \qquad (7)$$

The tensor energy density $\mathcal{H}_T$ now depends on six coupling constants, $C_t^{J0}$, $C_t^{J1}$, and $C_t^{J2}$, for $t$=0,1, and not on two coupling constants, $t_e$ and $t_o$, as in Eq. 3. Similarly, the SO energy density $\mathcal{H}_{SO}$ depends now on two coupling constants $C_t^{\nabla J}$ for $t$=0,1, and not on one, $W_0$, (the latter generalization has been introduced and studied in Ref.[17])

From the symmetry conditions imposed by the spherical, axial, and reflection symmetries one obtains[18] that:

1° The pseudoscalar densities $\underline{J}_t$ vanish unless the axial or reflection symmetries are broken.

2° The pseudotensor densities $\underline{\mathsf{J}}_t$ vanish unless the spherical symmetry is broken.

Finally, the gauge-invariance symmetry conditions[4] require that there are only two gauge-invariant combinations of the pseudoscalar, vector, and pseudotensor terms, namely,

$$\begin{aligned}
G_t^T &= \tfrac{1}{3}\underline{J}_t^2 + \tfrac{1}{2}\boldsymbol{J}_t^2 + \underline{\mathsf{J}}_t^2,\\
G_t^F &= \tfrac{2}{3}\underline{J}_t^2 - \tfrac{1}{4}\boldsymbol{J}_t^2 + \tfrac{1}{2}\underline{\mathsf{J}}_t^2.
\end{aligned} \qquad (8)$$

In such a case, only four out of the six tensor coupling constants are linearly independent, i.e.,

$$\begin{aligned}
C_t^{J0} &= \tfrac{1}{3} A_t + \tfrac{2}{3} B_t,\\
C_t^{J1} &= \tfrac{1}{2} A_t - \tfrac{1}{4} B_t,\\
C_t^{J2} &= A_t + \tfrac{1}{2} B_t.
\end{aligned} \qquad (9)$$

On the other hand, the averaging of the tensor forces (1) implies that only two out of the six tensor coupling constants remain linearly independent, i.e.,

$$B_0 = -3A_0 = -\frac{3}{8}\left(t_e + 3t_o\right),$$
$$B_1 = -3A_1 = \;\; \frac{3}{8}\left(t_e - t_o\right). \tag{10}$$

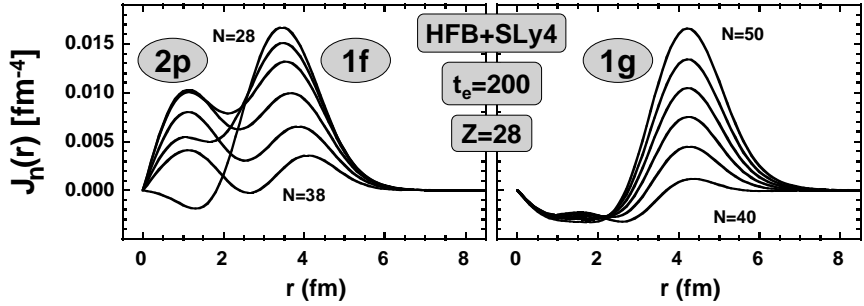

Figure 2.   Radial components $J_n(r)$ of the neutron vector SO densities $\boldsymbol{J}_n$, calculated for the $Z$=28 isotopes with $N$=28–38 (left panel) and $N$=40–50 (right panel).

In order to illustrate the influence of tensor densities on single-particle and global nuclear properties, in Fig. 2 are shown the radial components, $J_n(r)$=$\boldsymbol{J}_n \cdot \boldsymbol{r}/r$, of the neutron vector SO densities $\boldsymbol{J}_n$, calculated for the nickel isotopes between $N$=28 and 50. Calculations have been performed for the Skyrme SLy4 interaction,[19] by using the Hartree-Fock-Bogolyubov (HFB) method with the spherical symmetry assumed.[20]

One can see that the SO densities are mostly positive and peaked near the surface. At $N$=28, the SO density is large and its major part comes from the occupied $\nu f_{7/2}$ orbital (SUS system). By adding neutrons in the shell above $N$=28, this part is gradually cancelled by an increasing in magnitude, negative contribution from the SO partner $\nu f_{5/2}$. At the same time contributions from the $\nu p_{3/2}$ and $\nu p_{1/2}$ orbitals appear. When both pairs of the SO partners are occupied around $N$=38, and when the $\nu g_{9/2}$ orbital is still empty (SS system), the SO density is rather small. Beyond $N$=40, it increases again until the $\nu g_{9/2}$ orbital becomes fully occupied at $N$=50. Note the shift of the SO densities to larger distances, which occurs at the point of the switch-over between the dominating $1f$ and $1g$ contributions.

A similar pattern of varying SO densities is valid for all shells. For SS systems, one obtains small SO densities, while for SUS systems, the SO densities are large. Therefore, the SO densities are small at magic shells $N, Z=2$, 8, and 20, and large at magic shells $N, Z=28$, 50, 82, and 126.

Since the $j_>$ partners are always occupied first, the SO densities are mostly positive. Note that the derivatives of particle densities are mostly negative, and also peaked at the surface; therefore, for positive coupling constants, the SO and tensor forces split the SO partners in opposite directions, cf. Eq. (4).

### 4.2. *Single-particle levels*

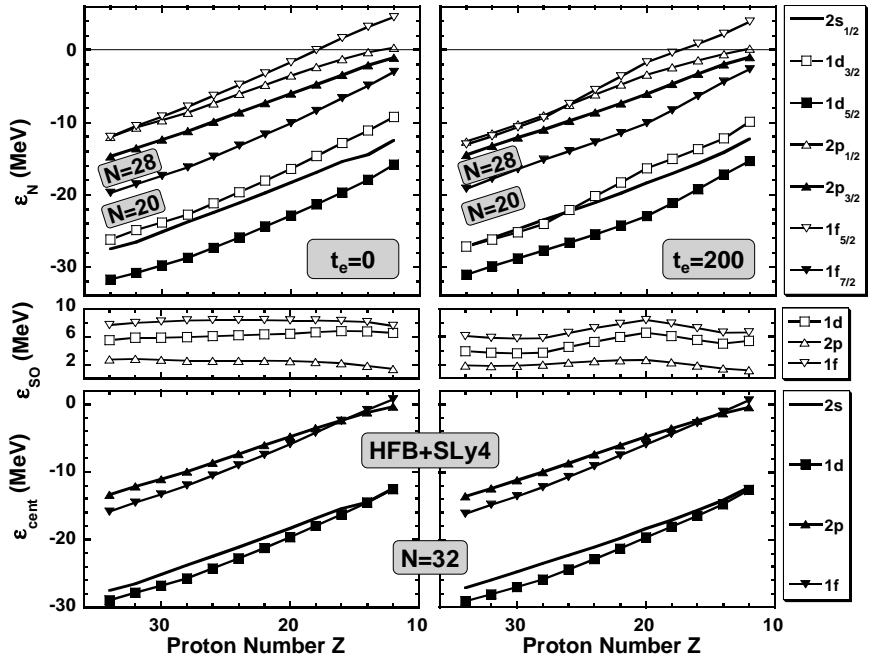

Figure 3. Single-particle properties of neutron levels in $N=32$ isotones, calculated within the HFB method with the SLy4 Skyrme interaction. Left panels correspond to the standard SLy4 parametrization with no tensor terms ($t_e=0$) while the right panels correspond to the tensor-even interaction included ($t_e=200$ MeV fm$^5$). Single-particle energies in the *sdpf* shell (top panels) are shown along with their spin-orbit-splitting energies (middle panels) and centroid energies (bottom panels).

In Fig. 3 are shown properties of neutron single-particle levels calculated for the chain of the $N=32$ isotones (these levels are relevant for the changes of the shell structure discussed in Sec. 2). Single-particle energies $\varepsilon_N^{n\ell j}$ (top panels) were calculated as the canonical energies[21] of the HFB method. In order to better visualize the dependence of the single-particle energies on the proton number, in the middle and bottom panels are shown the SO splittings and centroids, respectively, of the SO partners, defined as

$$
\begin{aligned}
\varepsilon_{\text{SO}}^{n\ell} &= \varepsilon_N^{n\ell j<} - \varepsilon_N^{n\ell j>}, \\
\varepsilon_{\text{cent}}^{n\ell} &= \tfrac{1}{2}\left(\varepsilon_N^{n\ell j<} + \varepsilon_N^{n\ell j>}\right).
\end{aligned}
\tag{11}
$$

Without the tensor terms (left panels of Fig. 3), the neutron single-particle energies vary smoothly with the proton numbers. Apart from a gradual increase with decreasing $Z$, one observes two clear type of changes in the shell structure. First, in each shell the centroids of levels with different values of $\ell$ become degenerate towards the neutron drip line. This effect is related to the increase of the surface diffuseness of particle distributions, which renders the shell structure of very neutron-rich nuclei similar to that of a harmonic oscillator.[22] Second, near the neutron drip line the SO splitting of the weakly bound $p$ orbitals becomes smaller, because such orbitals start to decouple from the SO potential due to their increasing spatial dimensions.[23]

In the right panels of Fig. 3 are shown the analogous results obtained with the tensor-even interaction $\hat{V}_{Te}$, Eq. (1), taken into account for $t_e=200\,\text{MeV}\,\text{fm}^5$. (This particular value of the coupling constant was not optimized in any sense, and it is used here only to illustrate some general trends.) The use of the tensor-even interaction ($T=0$, $S=1$ neutron-proton channel) corresponds to the energy density that depends on the product of neutron and proton SO densities, cf. Eq. (3). Therefore, the effect of the tensor term vanishes at the closed proton shell $Z=20$ (SS system). For $Z$ higher (lower) than 20, the effect of the tensor term increases as a result of increasing contributions to the proton SO density coming from the $\pi f_{7/2}$ ($\pi d_{3/2}$) orbitals. This is clearly visible in the middle right panel of Fig. 3, where the SO splitting decreases on both sides of $^{52}$Ca. This is so because, for positive values of the coupling constant $t_e$, the effect of the tensor force partly cancels that of the standard SO force. As a result, the $\pi f_{5/2}$ orbital is in $^{60}$Ni much closer to the $p$ orbitals than it is in $^{54}$Ti; the shift which is compatible with the changes of the shell structure discussed in Sec. 2.

A detailed reproduction of the level positions is not the goal of the present study. The coupling constants of the tensor terms have to be adjusted together with other parameters of the EDF, by considering not only this particular region of nuclei, and not only this particular set of observables. Indeed, the tensor terms included in the EDF will influence many different global nuclear properties throughout the mass chart, and a global analysis is therefore necessary. Before this is done, in the next section, the impact of the tensor interaction on nuclear binding energies is studied in a preliminary way.

### 4.3. *Binding energies*

When the tensor terms (3) are added to the EDF, the binding energies are affected through self-consistent changes of all the terms in the EDF. However, qualitatively, the effects of tensor terms on the ground-state energies can be illustrated by integrals of products of the SO densities that appear in Eq. (3). In Fig. 4, values of such integrals are shown for the neutron SO densities squared, $J_n^2(r)$, calculated at magic proton numbers in function of the neutron numbers. Comparison of results obtained without (left panel) and with (right panel) tensor-even interaction included, shows that the effect of the tensor term can, in the first approximation, be treated perturbatively. Due to the fact that the proton SO densities depend weakly on the neutron numbers, for $Z$=28, 50, and 82 the integrals of products $J_n(r)J_p(r)$ show similar a behaviour to those of $J_n^2(r)$, while they are small for $Z$=8 and 20.

From the results shown in Fig. 4, it is clear that, for positive coupling constants, the tensor terms will give characteristic contributions to the ground-state energies of heavy nuclei. These contributions will have a form of inverted arches, spanned between the neutron magic numbers. This feature is conspicuously reminiscent of differences between the theoretical and experimental ground-state energies obtained with tensor terms *not* included.[24,25,26] It is therefore quite plausible that by including the tensor terms one may be able to remove a major part of the discrepancy between the previously calculated nuclear masses and experiment.

## 5. Conclusions

During the past thirty odd years, when the nuclear self-consistent mean-field methods based on effective interactions were developed and implemented, the tensor interactions have been largely ignored. On the other hand, recent experimental studies and shell-model analyses indicate that these interactions may play an important role in several regions of nuclear chart. A unified picture of the role played by tensor interactions through-out the mass table is still missing, and is very much needed.  In the present paper, I have reviewed basic properties of the tensor mean fields, and I have illustrated their role in changing the shell structure and masses of nuclei.

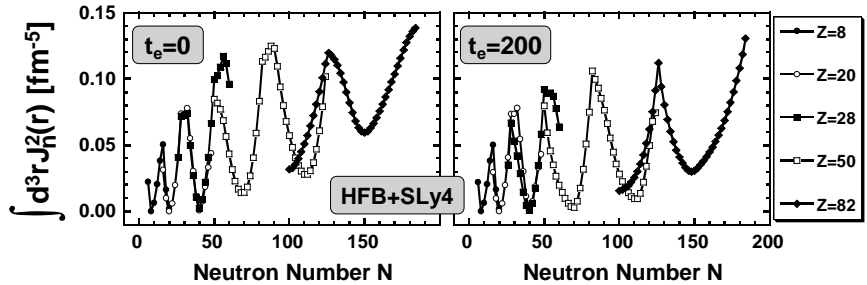

Figure 4.   Integrals of the neutron SO densities squared that define contributions of the tensor terms to total binding energies.  Left panels correspond to the standard SLy4 parametrization with no tensor terms ($t_e$=0), while the right panels correspond to the tensor-even interaction included ($t_e$=200 MeV fm$^5$).

## Acknowledgments

This work was supported in part by the Polish Committee for Scientific Research (KBN) under contract N0. 1 P03B 059 27 and by the Foundation for Polish Science (FNP).